\documentclass[final,5p,times,twocolumn]{elsarticle}
\usepackage{amssymb}
\usepackage{amsmath}
\usepackage[mathscr]{euscript} 
\usepackage[]{hyperref} 

\usepackage{color}

\journal{Journal of Magnetism and Magnetic Materials}

\begin{document}

\begin{frontmatter}


\title{Suppression of the skyrmion Hall effect in planar nanomagnets by the magnetic properties engineering: Skyrmion transport on nanotracks with magnetic strips}
%

%

%
%
%


\author[address1]{D. Toscano\corref{correspondingauthor}}
\cortext[correspondingauthor]{Corresponding author: Danilo Toscano}
\ead{dtoscano@ice.ufjf.br}

\author[address1]{J.P.A. Mendon\c{c}a}

\author[address1]{A.L.S. Miranda}

%

\author[address2]{C.I.L. de Araujo}

\author[address1]{F. Sato}

\author[address1]{P.Z. Coura}

\author[address1]{S.A. Leonel}



\address[address1]{Departamento de F\'{\i}sica, Laborat\'orio de Simula\c{c}\~ao Computacional, Universidade Federal de Juiz de Fora, Juiz de Fora, Minas Gerais 36036-330, Brazil}

\address[address2]{Departamento de F\'{\i}sica, Laborat\'orio de Spintr\^onica e Nanomagnetismo, Universidade Federal de Vi\c{c}osa, Vi\c{c}osa, Minas Gerais 36570-900, Brazil}

%

\begin{abstract}
 
\noindent Micromagnetic simulations have been performed to investigate the suppression of the skyrmion Hall effect in  nanotracks with their magnetic properties strategically modified. In particular, we study two categories of magnetically modified nanotracks. One of them, repulsive edges have been inserted in the nanotrack and, in the other, an attractive strip has been placed exactly on the longest axis of the nanotrack. Attractive and repulsive interactions can be generated from the engineering of magnetic properties. For instance, it is known that the skyrmion can be attracted to a region where the exchange stiffness constant is decreased. On the other hand, the skyrmion can be repelled from a region characterized by a local increase in the exchange stiffness constant. In order to provide a background for experimental studies, we vary not only the magnetic material parameters (exchange stiffness, perpendicular magnetocrystalline anisotropy and the Dzyaloshinskii-Moriya constant) but also the width of the region magnetically modified, containing either a local reduction or a local increase for each one of these magnetic properties. In the numerical simulations, the skyrmion motion was induced by a spin-polarized current and the found results indicate that it is possible to transport skyrmions around the longest axis of the nanotrack. In practice, the skyrmion Hall effect can be completely suppressed in magnetic nanotracks with strategically modified magnetic properties. Furthermore, we discuss in detail 6 ways to suppress the skyrmion Hall effect by the usage of nanotracks with repulsive edges and nanotracks with an attractive strip.

\end{abstract}



\end{frontmatter}


\section{Introduction}
\label{intro}

Magnetic skyrmions are nanoscaled topological spin textures, which behave as quasiparticles and have high potential of being information
carriers in novel spintronic technologies~\cite{JPhysD_ApplPhys_44_392001_2011,NatureNanoTechnology_8_899_2013,NatureNanoTechnology_8_152_2013}. In the beginning, skyrmions have been experimentally observed only at low temperatures and under the influence of large external magnetic fields~\cite{Science_323_915_2009,PhysRevB_81_041203_2010,Nature_465_901_2010,NaturePhysics_7_713_2011,Nat_Mater_10_106_2011,
Science_341_636_2013}. Nowadays, these quasiparticles have been stabilized at room temperature in magnetic multilayer systems with interfacial Dzyaloshinskii-Moriya couplings and high perpendicular magnetic anisotropy~\cite{Science_349_283_2015,ApplPhysLett_106_242404_2015,Nat_Mater_15_501_2016,NatureNanoTechnology_11_444_2016,NatureNanoTechnology_11_449_2016,
ApplPhysLett_111_202403_2017,ApplPhysLett_112_132405_2018}. 
The stability, nucleation and annihilation of skyrmions have been extensively investigated~\cite{Science_349_283_2015,NatureNanoTechnology_8_839_2013,PhysRevB_85_174416_2012,ApplPhysLett_102_222405_2013,PhysRevB_88_184422_2013,NatureNanoTechnology_8_742_2013,Nat_Commun_5_4652_2014,
PhysRevLett_114_177203_2015,JPhysD_ApplPhys_48_115004_2015,PhysRevLett_110_167201_2013,Scientific_Reports_5_17137_2015,
AIP_Advances_5_047141_2015,PhysRevB_93_024415_2016}.

Understanding and controlling of skyrmion motion in magnetic nanowires is extremely important for the development and realization of spintronic devices~\cite{Nat_Mater_6_813_2007,JPhysD_ApplPhys_49_423001_2016,Nat_Rev_Mater_2_17031_2017}. Due to the peculiar characteristics of skyrmions, such as nanoscaled sizes, topological protection and efficient electric manipulation, there is a lot of interest in replacing domain walls with skyrmions in the famous racetrack memory~\cite{DW_RacetrackM_2008} and others spintronic technologies~\cite{NatureNanoTechnology_8_152_2013}. The skyrmion transport in nanotracks~\cite{Scientific_Reports_4_6784_2014} can be driven by the spin-transfer torque (STT) or the spin-Hall effect (SHE).
As highlighted in Ref.~\cite{NaturePhysics_13_112_2017}, the skyrmion Hall effect~\cite{NatureNanoTechnology_8_899_2013} has been experimentally observed by two groups simultaneously~\cite{NaturePhysics_13_162_2017,NaturePhysics_13_170_2017}. Due to their topological charge, ferromagnetic skyrmions can not be driven by spin-polarized currents without being moved away from the longitudinal axis of the nanotrack. From the technological point of view, the skyrmion Hall effect is an issue to the development of skyrmion-based racetrack memories~\cite{NatureNanoTechnology_8_839_2013}. Several strategies have been proposed to suppress this undesirable phenomenon. For instance, the spatial variations in the nanotrack thickness, by inserting thicker edges than the central region~\cite{Scientific_Reports_5_10620_2015} or a groove (thinner central region) along the longest axis of the nanotrack~\cite{IEEE_51_1500204_2015}. However, the incorporation of  these non-magnetic defects patterned as part of the nanotrack are accompanied by the spatial variation of the magnetic material parameters, for example, perpendicular magnetic anisotropy (PMA) and strength Dzyaloshinskii-Moriya interaction (DMI). Instead of modifying the nanotrack thickness, the reference~\cite{Scientific_Reports_7_45330_2017} proposes the incorporation of a material with high perpendicular magnetocrystalline anisotropy at the two symmetrical edges of the nanotrack. An alternative method would be to use ion irradiation to modify the magnetic properties in a selected area of the nanotrack~\cite{Ion_Irradiation_1998,Review_Implantation_2008}. Besides providing more parameters to control the strength of the pinning (potential wells) or blocking (potential barriers) regions, magnetic defects consisting in spatial variations on the material parameters are more interesting because they do not distort the nanotrack geometry. Fook et al.~\cite{IEEE_51_1500204_2015} observed the confinement of skyrmions along of a strip with reduced anisotropy located on the axis of the nanotrack. The authors suggested the irradiation of He$^{+}$ ion in a Co/Pt multilayer to reduce the perpendicular anisotropy of the irradiated area, a practice which is known since 1998~\cite{Ion_Irradiation_1998}. Effects of spatially engineered Dzyaloshinskii-Moriya interaction have already been investigated in the literature~\cite{PhysRevB_95_144401_2017,PhysRevB_96_214403_2017}, and it was reported that skyrmions can be confined within a region with higher DM interaction. 

\begin{figure}[htb!]
\centering
	\includegraphics[width=8.0cm]{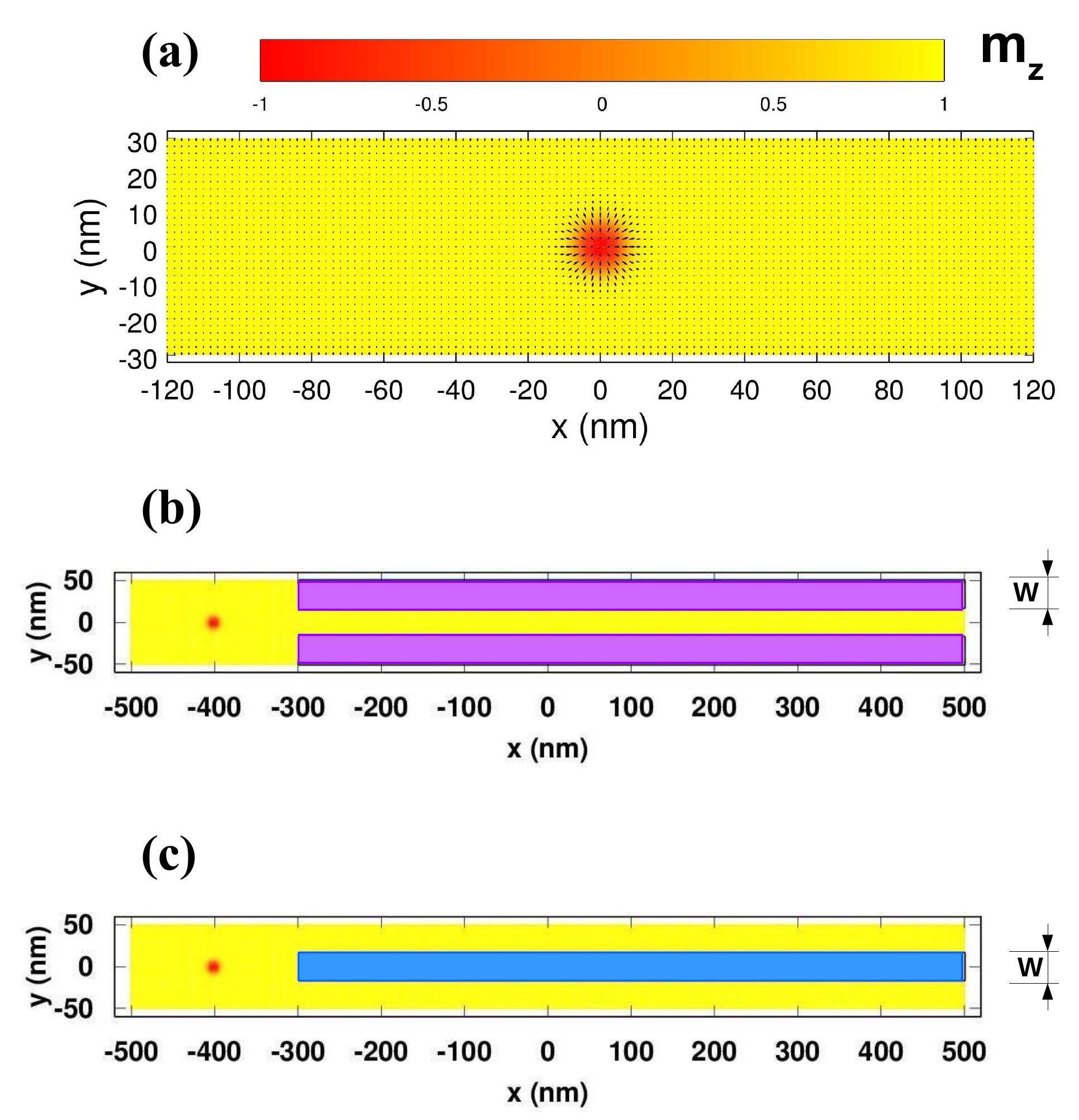}
\caption{(Color online). Schematic views of a rectangular nanomagnet, hosting a single magnetic skyrmion.  Fig (a) highlights that  the majority of the nanowire's magnetic moments is going out of the plane of the figure except at the core of the skyrmion (red region), where they are pointing in the opposite direction. For simplicity, the arrows representing magnetic moments are not displayed in figures (b) and (c), which show two categories of nanotracks modified magnetically. The blue and pink regions consist in spatial variations of the material parameters, which have been intentionally modified in order to confine the skyrmion along the center region of the nanotrack. Attractive and repulsive effects can be achieved when tuning either a local increase ($X''>X$) or a local reduction ($X''<X$), where $X$ is a target magnetic property: exchange stiffness $A$, perpendicular magnetocrystalline anisotropy $K$ or the Dzyaloshinskii-Moriya constant $D$. Fig (b) shows a nanotrack with repulsive edges ($A''>A$ or $D''<D$ or $K''>K$), whereas in the Fig (c) shows a  nanotrack with an attractive strip ($A''<A$ or $D''>D$ or $K''<K$). The width $W$ of the region modified magnetically is a crucial parameter to suppress the skyrmion Hall effect.}
\label{fig:Schematic}
\end{figure}

Although the issue of the skyrmion Hall effect is automatically suppressed in an antiferromagnetic medium~\cite{Scientific_Reports_6_24795_2016,PhysRevLett_116_147203_2016,ApplPhysLett_109_182404_2016} (as a consequence of the net topological charge to be identically zero), in this work we investigate the suppression of the skyrmion Hall effect in ferromagnetic nanotracks with their magnetic properties strategically modified along the stripes parallel to the longest axis of the nanotrack. Our setup is shown in Fig. (\ref{fig:Schematic}). It is worth mentioning that the present work explores the results of a previous work~\cite{JMagnMagnMater_480_171_185_2019}, where our team has reported that spatial variations on the material parameters of the ferromagnetic medium (exchange stiffness, saturation magnetization, magnetocrystalline anisotropy and Dzyaloshinskii-Moriya constants) can work as traps for pinning and scattering magnetic skyrmions.


%
\section{Model and Methodology}
\label{cidwm}

In order to describe the ferromagnetic nanotrack, we have considered exchange and Dzyaloshinskii-Moriya interactions, perpendicular magnetic anisotropy, and shape anisotropy, included in the following Hamiltonian:
%
%
%
%
%
%
%
\begin{eqnarray}
\mathscr{H}  & = & \:- \:\sum_{<i,j>}\: J_{ij}\: \left [\:\hat{m}_{i}\cdot\hat{m}_{j}\:\right ] \:\:+{}                                        
                    \nonumber \\
                    \nonumber \\
&& {}  -  \sum_{<i,j>}\: D_{ij}\: \left [ \:\hat{d}_{ij}\cdot (\hat{m}_{i}\times \hat{m}_{j})\: \right ] \:\:+{}
                    \nonumber \\
                    \nonumber \\
&& {}  - \:\sum_{i}\:  K_{i} \left [\: \hat{m}_{i}\cdot\hat{n}\: \right ]^{2} \:\: + \:\:\mathscr{H}_{\textrm{dip}} 
 \label{hamiltonian}
\end{eqnarray}
where $ \hat m_{k} \equiv (m^{x}_{k},m^{y}_{k},m^{z}_{k})$ is a dimensionless vector, corresponding to the magnetic moment located at the site $k$ of the lattice.
Due to the short range of the exchange interaction, the summation is over the nearest magnetic moment pairs $<i,j>$. The second term in Eq. (\ref{hamiltonian}) represents the Dzyaloshinskii-Moriya interactions, where the versor $\hat{d}_{ij}$ depends on the type of magnetic system considered. For a magnetic multilayer system presenting the interfacial DMI, the versor $\hat{d}_{ij}=\hat{u}_{ij}\times \hat{z}$, where $\hat{z}$  is a versor perpendicular to the multilayer surface and $\hat{u}_{ij}$ is unit vector joining the sites $i$ and $j$ in the same layer~\cite{PhysRevB_88_184422_2013,PhysRevLett_115_267210_2015}. These magnetic systems favor the nucleation of  N\'{e}el skyrmions (hedgehog-type configuration). The third term in Eq. (\ref{hamiltonian}) describes the uniaxial magnetocrystalline anisotropy, since $K_{i}>0$ and $\hat{n}=\hat{z}$. 
The strength of the magnetic interactions, $J_{ij}, \: D_{ij}, \: K_{i}$ 
have the same dimension (energy unity) and they assume different values depending on the spatial variation of the  material parameters: exchange stiffness constant $A$, Dzyaloshinskii-Moriya constant $D$, magnetocrystalline anisotropy constant $K$  and saturation magnetization constant $M_{\mbox{\tiny{S}}}$. In our simulations we use the micromagnetic approach, in which the work cell has an effective magnetic moment $\vec{m}_{k}=\left(M_{\mbox{\tiny{S}}}\:V_{cell}\right)\:\hat{m}_{k}$. For the case in which the magnetic system is discretized into cubic cells $V_{cell}=a^{3}$, the possible values for the strength of the magnetic interactions are given by:
\begin{equation}
J_{ij}= 2\: a\: \left\{\begin{array}{l}
         A \\
         A' \\
         A''   
             \end{array}\right.
\label{eq:A}
\end{equation}

\begin{equation}
D_{ij}= a^{2} \left\{\begin{array}{l}
         D \\
         D' \\
         D''   
             \end{array}\right.
\label{eq:D}
\end{equation}

\begin{equation}
K_{i}= a^{3} \left\{\begin{array}{l}
         K \\
         K''   
             \end{array}\right.
\label{eq:K}
\end{equation}
%
%
where $A, D, K$ are parameters of the host material, and $A'', D'', K''$ are the parameters of the guest material. $A', D'$ are parameters which describe interactions at the interface between two ferromagnetic materials. The geometric mean was adopted for the interface parameters: $A'=\sqrt{A\:\cdot\:A''\:\:}$ and $D'=\sqrt{D\:\cdot\:D''\:\:}$ in order to allow the magnetic parameters to vary gradually from a magnetic medium to the other.
%
%
%

The magnetization dynamics is governed by the Landau-Lifshitz-Gilbert (LLG) equation~\cite{Landau_Lifshitz,Gilbert}, which includes the Zhang-Li spin transfer torque~\cite{zhang2004}:
\begin{eqnarray}
\frac{\partial \hat m_i}{\partial t'} = - \frac{1}{(1+\alpha^2)} \biggl \{ \hat m_i \times \vec b_i + \alpha \:\hat m_i \times \left( \hat m_i \times \vec b_i \right )+ \nonumber \\ \nonumber \\ \nonumber  + \frac{1}{(1+\beta^2)} \biggl ( \frac{u}{a\: \omega_0} \biggr ) \biggl [ \left (\beta -\alpha \right )\: \hat m_i \times \frac{\partial \hat m_i}{\partial x'} \:+\\\nonumber  \\+  \left(1+\alpha \beta\right ) \: \hat m_i \times \left( \hat m_i \times \frac{\partial  \hat m_i}{\partial  x'}\right ) \biggl ] \biggr \}                                                                                        
 \label{eq_LLG}
\end{eqnarray}
where $\vec{b}_{i}=-\left( \frac{1}{2\:a\:A}\right)\frac{\partial \mathscr{H}}{\partial \hat{m}_{i}}$ is the dimensionless effective field at the lattice site $i$. 
The connection between the space-time coordinates and their dimensionless corresponding is given by: $\Delta x' = \Delta x / a$ and $\Delta t' = \omega_{0} \: \Delta t$, where $\omega_0 = (\frac{\lambda}{a})^2 \mu_0 \gamma M_s$ is a  scale factor with inverse time dimension, being $\gamma$ the electron gyromagnetic ratio and $\lambda=\sqrt{\frac{2A}{\mu_0 M_{\mbox{\tiny{S}}}^2}}$ the exchange length. The product $(a\: \omega_0)$ has the dimension of distance divided by time (unit of velocity) as well as the term $u = j_e \left(\frac{\mu_{\mbox{\tiny{B}}} P }{e M_s}\right) $, where $\mu_{\mbox{\tiny{B}}}$ is the Bohr magneton, $j_e$ is a component of the electric current density vector. In our simulations we apply electrical current density $\vec{j}_{e}=-\vert j_{e}\vert \:\hat{x}$ (conventional current) in order to move the skyrmion from the left to the right, see Figs. \ref{fig:Schematic} (b) and \ref{fig:Schematic}(c) that show our initial conditions. The LLG explicit equation, see equation (\ref{eq_LLG}), was integrated by using a fourth-order predictor-corrector scheme with time step $\Delta t'=0.01$. Before the spin-polarized current to be applied, we performed relaxation micromagnetic simulations in order to obtain a single skyrmion at the position $(X_{\mbox{\tiny{S}}},\:Y_{\mbox{\tiny{S}}})=(-\:400, \: 0)\:\: \textrm{nm}$.  More specifically, we have used as initial condition of the LLG equation without external agent (magnetic field or spin-polarized current) an approximated magnetization configuration. By solving the LLG equation, we can lead the magnetic system
to the minimum energy configuration. This makes possible the adjustment of the skyrmion diameter. The equilibrium configurations obtained in this way have been used as initial configuration in other simulations, in which a spin-polarized current was applied in order to move the skyrmion for the region of the magnetic defect. In the analytical solution, a single skyrmion with radius $R_{\mbox{\tiny{S}}}$ is placed exactly into the intended location $(X_{\mbox{\tiny{S}}},Y_{\mbox{\tiny{S}}})$. When using two scalar fields in a cylindrical coordinate system, the initial configuration representing the skyrmion can be written as
\begin{equation}
\hat m_i=(\cos{\Theta_{i}}\cos{\Phi_{i}}\:,\:\cos{\Theta_{i}}\sin{\Phi_{i}}\:,\:P_{\mbox{\tiny{S}}}\sin{\Theta_{i}})
\end{equation}
where the scalar fields are given by~\cite{Bazeia_2016,JAP_119_193903_2016}

\begin{equation}
	\Theta_{i}=\frac{\pi}{2}\left[\frac{1-\left(\frac{R_{i}}{R_{\mbox{\tiny{S}}}}\right)^{\frac{2}{1-s}}}{1+\left(\frac{R_{i}}{R_{\mbox{\tiny{S}}}}\right)^{\frac{2}{1-s}}}\right]
	\label{eq_theta}
\end{equation}

\begin{equation}
	\Phi_{i}=q\:\phi_{i}\:+\varphi
\label{eq_phi}	
\end{equation}
\vspace{0.25cm}

\noindent where $R_{i}=\sqrt{\left(x_{i}-X_{\mbox{\tiny{S}}}\right)^{2}+\left(y_{i}-Y_{\mbox{\tiny{S}}}\right)^{2}\:}$ and $\phi_{i}=\arctan{\left( y_{i}/x_{i} \right)}$ are cylindrical coordinates of the position vector $\vec{r}_{i}=(x_{i},\:y_{i},\:z_{i})=\left(R_{i}\cos{\phi_{i}},\:R_{i}\sin{\phi_{i}},\:z_{i}\right)$. At the core of the magnetic skyrmion, the out-of-plane magnetization can point either up $(P_{\mbox{\tiny{S}}}=+1)$ or down $(P_{\mbox{\tiny{S}}}=-1)$. This structural property can be referred to as the \textit{polarity} of the skyrmion, being that away from the skyrmion core the out-of-plane magnetization points in the opposite direction of $P_{\mbox{\tiny{S}}}$. The $s$-parameter is in the interval $0 \leq s < 1$. The phase constant $\varphi = n\:\dfrac{\pi}{2}$ is related to the type and the \textit{chirality} of the skyrmion $C_{\mbox{\tiny{S}}}$, specifically:
\begin{equation}
	n=    \left\{\begin{array}{l}
		1, ~\scriptstyle \textrm{Bloch skyrmion with~} C_{\mbox{\tiny{S}}} \:=\: +1~ (\textrm{anticlockwise rotation}) \\
		2, ~\scriptstyle \textrm{N\'{e}el skyrmion with~} C_{\mbox{\tiny{S}}} \:=\: -1~ (\textrm{inward radial flow})\\
		3, ~\scriptstyle \textrm{Bloch skyrmion with~} C_{\mbox{\tiny{S}}} \:=\: -1~ (\textrm{clockwise rotation}) \\
		4, ~\scriptstyle \textrm{N\'{e}el skyrmion with~} C_{\mbox{\tiny{S}}} \:=\: +1~ (\textrm{outward radial flow})   
	\end{array}\right.
	\label{eq:C}
\end{equation}    
The \textit{topological vorticity} $q$ of the quasiparticle is a quantized quantity. For particles (vortices and skyrmions) $q\:=\:+1,+2,+3,\dots$, whereas for anti-particles (anti-vortices and anti-skyrmions) $q\:=\:-1,-2,-3,\dots$ It is important do not confuse topological vorticity~\cite{wysin_2010,JAP_109_014301_2011}

\begin{equation}
q=\frac{1}{2\pi}\oint_{C} \:\vec{\nabla} \phi\cdot d\vec{l}  
\end{equation}
with topological charge~\cite{NatureNanoTechnology_8_899_2013,NaturePhysics_13_162_2017}

\begin{equation}
Q=\frac{1}{4\pi}\iint\limits_{S} \hat m \cdot \left( \frac{\partial  \hat m}{\partial  x} \times \frac{\partial  \hat m}{\partial  y}\right )\:dx\:dy
\end{equation}
which is also referred to as the skyrmion topological number $N_{\mbox{\tiny{sk}}}$. In the literature, this confusion occurs because we have $Q=q=1$ for skyrmions.
\noindent For vortices $Q=N_{\mbox{\tiny{sk}}}=\pm 1/2$, whereas $Q=N_{\mbox{\tiny{sk}}}=\pm 1$ for skyrmions.
The N\'{e}el skyrmion with outward radial flow and polarity downward has been used in our simulations, see Fig. \ref{fig:Schematic} (a). Such skyrmion is characterized by a negative topological charge and we obtain numerically the value $Q = -0.9303 \approx -1$, which is agreement with the theoretical value. When applying a spin-polarized current (Zhang-Li spin-transfer torques), any skyrmion profile will be shifted in the direction of the electron flow. However, the transversal component of the motion depends on the sign of the topological charge and on the ratio ($\beta/\alpha$). The influence of the ratio ($\beta/\alpha$) on the skyrmion Hall angle has already been investigated in Refs.~\cite{NatureNanoTechnology_8_839_2013,NatureNanoTechnology_8_742_2013}. For $\beta >\alpha$ and the current flowing from the left to the right, a skyrmion with negative topological charge moves upward ($Y_{\mbox{\tiny{S}}}>0$) in our setup, see Fig. (\ref{fig:Schematic}). The schematic of the skyrmion Hall effect can be understood using a Thiele approach~\cite{PhysRevLett_30_230_1973} and it is very-well discussed in Ref.~\cite{NaturePhysics_13_162_2017}.

For the geometric parameters of the nanotracks, we have considered the length $L_{x}=1000\:\textrm{nm}$, the width $L_{y}=100\:\textrm{nm}$ and the thickness $L_{z}=2\:\textrm{nm}$, differing one to another only in the parameters of the magnetic strips. In the simulations we used the typical parameters for Co/Pt multilayers. In the Table (\ref{tab:parameters}) we present the material parameters used in micromagnetic simulations. At the end of the relaxation simulations, we have estimated the value of the skyrmion radius as $R_{\mbox{\tiny{S}}}\approx 8.3\:\textrm{nm}$.
%
%
Although we have not taken into account the full dipolar coupling, that is, $\mathscr{H}_{\textrm{dip}}=0$ in the equation (\ref{hamiltonian}), we took into account the shape anisotropy of a planar nanomagnet by considering an effective uniaxial anisotropy~\cite{PhysRevB_88_184422_2013}
\begin{equation}
K=K_{\mbox{\tiny{u}}}-\frac{\mu_{0}\:M_{\mbox{\tiny{S}}}^{2}}{2}
\label{eq:K_eff}
\end{equation}
Thus, the shape anisotropy that originates in the dipolar coupling was included in our simulations by considering an easy-plane anisotropy, that is,
\begin{equation}
\mathscr{H}_{\textrm{ shape}}^{\textrm{ anis}}\:=\:\left(\frac{\mu_{0}\:M_{\mbox{\tiny{S}}}^{2}}{2}\right)a^{3}\sum_{i}\:  \left [\: \hat{m}_{i}\cdot\hat{n}\: \right ]^{2} 
\end{equation}
where $\hat{n}=\hat{z}$ gives the direction of the magnetization hard axis.
For Co/Pt ultrathin films, we compute $K= 5.9 \times 10^{5}$ J/m$^{3}$. Unless otherwise stated, this value for the effective uniaxial anisotropy constant was used in most of our simulations. The \textcolor{blue}{supplementary material} describes carefully the Methodology used in this study, besides providing additional simulations that include or not the effect of the full dipolar coupling. There, we show that the effect of the full dipolar coupling is negligible for the simulated nanotrack. In other words, the results of the micromagnetic simulations remain qualitatively the same, provided the shape anisotropy be incorporated in the effective anisotropy and the skyrmion trajectory be distant from the boundary of the sample.


\begin{table}[b!]
\caption{Material parameters used in the micromagnetic simulations for Co/Pt nanotracks, see Refs.~\cite{NatureNanoTechnology_8_839_2013,Scientific_Reports_7_45330_2017}.}
\vspace{0.25cm}
\def\arraystretch{1.1}
\centering
\begin{tabular}{|c|c|}
%
\hline 
%
\small{exchange stiffness}               &        $A = 1.5 \times 10^{-11}$ J/m  \\
\hline 
\small{Dzyaloshinskii-Moriya constant}   &      $D = 3.0\times 10^{-3}$ J/m$^{2}$  \\
\hline 
\small{perpendicular anisotropy}         &      $K_{\mbox{\tiny{u}}} = 8.0 \times 10^{5}$ J/m$^{3}$  \\
\hline 
\small{saturation magnetization}         &     $M_{\mbox{\tiny{S}}}=5.8\times 10^{5}$ A/m  \\
\hline 
\small{work cell volume}                 &     $V_{\mbox{\tiny{cel}}}=a^{3}=\left(2 \times 2\times 2\right)$ nm$^{3}$  \\
\hline 
\small{damping parameter}                &     $\alpha=0.30$  \\
\hline 
\small{degree of non-adiabaticity}       &     $\beta=0.35$  \\
\hline 
\small{spin polarization of the current}       &     $P=0.7$  \\
\hline 
\small{electric current density}       &     $j_{e}=5.0\times 10^{12}$ A/m$^{2}$  \\
\hline 
\end{tabular}
\label{tab:parameters}
\end{table}

Spatial variations on the magnetic parameters were considered individually. 
%
Thus, we have studied three possible sources of magnetic defects: Type $A$, Type $D$ and Type $K$. Type $A$ magnetic defects are those characterized only by spatial variations in the exchange stiffness constant (other parameters of the magnetic material were unchanged in the region of the defect), Type $D$ magnetic defects are those characterized only by spatial variations in the Dzyaloshinskii-Moriya constant, and Type $K$ magnetic defects are those characterized only by spatial variations in the perpendicular magnetocrystalline anisotropy. We have considered magnetic defects in the shape of strips with different widths, where W ranging from 12 to 44 nm. Guest material parameters, $A'',\: D''$ and $K''$ were regarded as tuning parameters to obtain traps for the skyrmion, which contains either a local reduction or a local increase of a given magnetic property. 
%


In order to map the skyrmion position during the magnetization dynamics, we have used a Bioinspired Algorithm known as Frogs Method, which ``hunts'' the skyrmion by fitting the parameters of the equations (\ref{eq_theta}) and (\ref{eq_phi}). In this work, all trajectories have been obtained with 100 walkers and 100 convergence steps. The reference~\cite{mendona2019frogs} provides more details about the algorithm used to track the skyrmion position.

%
\section{Results and Discussion}
\label{Results}

In order to quantify the suppression of the skyrmion Hall effect, we have measured the value of the variable $Y^{*}_{\mbox{\tiny{S}}}$, which represents the average distance perpendicular to the longest axis of the nanotrack in which the skyrmion is transported inside the magnetically modified region. In the majority of simulations, we have observed
\begin{equation}
Y^{*}_{\mbox{\tiny{S}}}\:\approx\: Y_{\mbox{\tiny{S}}} \:(X_{\mbox{\tiny{S}}}=400\:\textrm{nm})
\end{equation}
As smaller value of $Y^{*}_{\mbox{\tiny{S}}}$ is, better is the suppression of the skyrmion Hall effect. In the reference nanotrack (without any magnetic strip), we measure the value of $Y^{*}_{\mbox{\tiny{S}}}=31.1\: \textrm{nm}$. Thus, the skyrmion Hall effect is completely suppressed as $Y^{*}_{\mbox{\tiny{S}}}\to 0$. 

From our previous work~\cite{JMagnMagnMater_480_171_185_2019}, we know that there are various sources of magnetic defects~\cite{JAP_109_076104_2011}, which can generate traps for ferromagnetic skyrmions, and multiple mechanisms can contribute simultaneously. In a scattering trap, the skyrmion moves away from the magnetic defect. This implies the existence of a repulsive interaction between the skyrmion and the magnetic defect, that is, a potential barrier. The notable repulsive effects can be achieved by three different ways: (i) Type $A$ magnetic defects characterized by a local increase in the exchange stiffness constant $A''>A$, (ii) Type $D$ magnetic defects characterized by a local reduction in the Dzyaloshinskii-Moriya constant $D''<D$ and (iii) Type $K$ magnetic defects characterized by a local increase in the perpendicular anisotropy constant $K''>K$. 

On the other hand, in a pinning trap the skyrmion moves towards the magnetic defect. This implies the existence of an attractive interaction between the skyrmion and the magnetic defect, that is, a potential well. The notable attractive effects can be achieved by three different ways: (i) Type $A$ magnetic defects characterized by a local reduction in the exchange stiffness constant $A''<A$, (ii) Type $D$ magnetic defects characterized by a local increase in the Dzyaloshinskii-Moriya constant $D''>D$ and (iii) Type $K$ magnetic defects characterized by a local reduction  in the perpendicular anisotropy constant $K''<K$.  

In Table (\ref{tab:traps}) we summarize the relevant informations needed to design nanotracks with their magnetic properties strategically modified.  

\begin{table}[htb!]
\caption{Six ways to suppress the skyrmion Hall effect by the usage of nanotracks with repulsive edges and nanotracks with an attractive strip, when tuning either a local increase ($X''\:>\:X$) or a local reduction ($X''\:<\:X$), where $X$ can be $\:A,\:\: D, \:\:\textrm{or}\:\: K $.}

\centering
\vspace{0.25cm}


\begin{tabular}{|c|c|}
%
\hline 
%
\textbf{Attractive Strip} &  \textbf{Repulsive Strip} \\
\hline
         $A'' <  A$   &        $A'' >  A$  \\
\hline 
         $D''\:>\:D $ &        $D''\: <\:D$ \\
\hline 
         $K'' \: <\: K$     &  $K''\:> \: K$  \\
\hline 
%
\end{tabular}

\label{tab:traps}
\end{table}

\subsection{\label{sec:edges}Nanotracks with repulsive edges}

\begin{figure}[b!]
\centering
	\includegraphics[width=7.7cm]{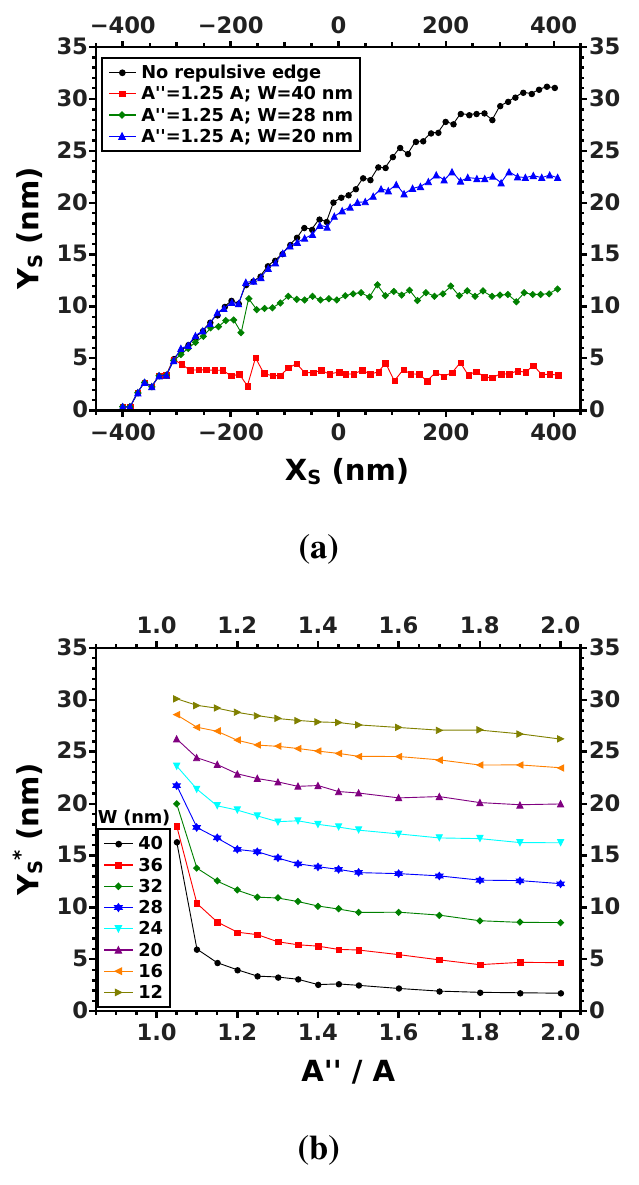}
\caption{(Color online). Spatial variations in the exchange stiffness constant: Type $A$ magnetic defects along the repulsive edge of width $W$. Figure (a) shows the skyrmion trajectory in nanotracks with different widths of the repulsive edges, presenting a local increase of 25\% in $A$. Figure (b) shows simulation results for $Y^{*}_{\mbox{\tiny{S}}}$, the variable that quantifies the suppression of the skyrmion Hall effect.}
\label{fig:A_edges}
\end{figure}

\begin{figure}[htb!]
\centering
	\includegraphics[width=7.7cm]{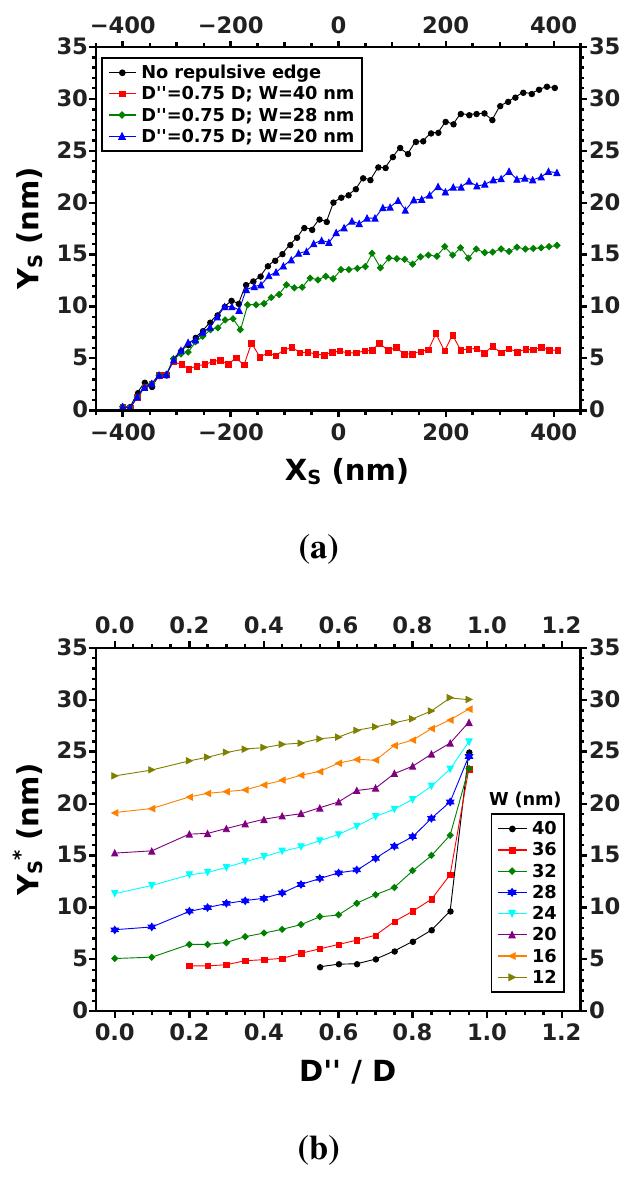}
\caption{(Color online). Spatial variations in the constant of the Dzyaloshinskii-Moriya interaction: Type $D$ magnetic defects along the repulsive edge of width $W$. Figure (a) shows the skyrmion trajectory in nanotracks with different widths of the repulsive edges, presenting a local reduction of 25\% in $D$. Figure (b) shows simulation results for $Y^{*}_{\mbox{\tiny{S}}}$, the variable that quantifies the suppression of the skyrmion Hall effect.}
\label{fig:DM_edges}
\end{figure}

\begin{figure}[htb!]
\centering
	\includegraphics[width=7.7cm]{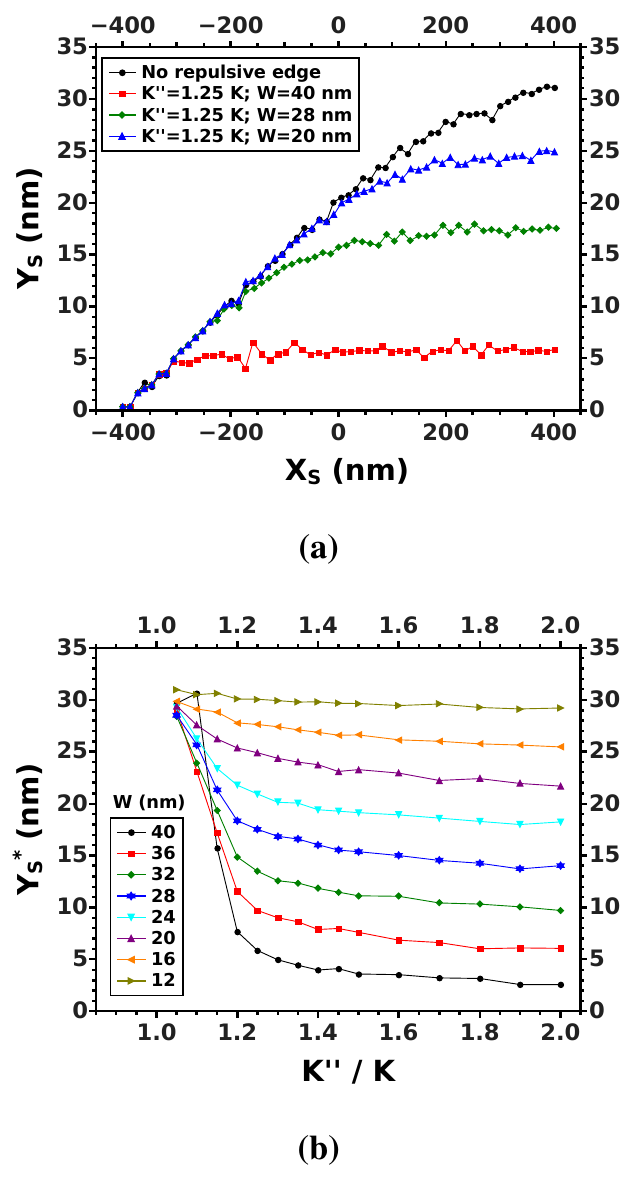}
\caption{(Color online). Spatial variations in the perpendicular magnetic anisotropy: Type $K$ magnetic defects along the repulsive edge of width $W$.  Figure (a) shows the skyrmion trajectory in nanotracks with different widths of the repulsive edges, presenting a local increase of 25\% in $K$. Figure (b) shows simulation results for $Y^{*}_{\mbox{\tiny{S}}}$, the variable that quantifies the suppression of the skyrmion Hall effect.}
\label{fig:K_edges}
\end{figure}

In this section, we investigate the suppression of the skyrmion Hall effect in nanotracks with repulsive edges, see Fig. \ref{fig:Schematic}(b). The effect of the repulsive edge on the skyrmion trajectory is shown in Figs. \ref{fig:A_edges}(a),\ref{fig:DM_edges}(a) and \ref{fig:K_edges}(a). 

As a reference, we plot the skyrmion trajectory in the corresponding nanotrack without repulsive edges. From these figures, one can see that the strength of the repulsive potential becomes stronger as the edge width $W$ increases. In order to systematically investigate the suppression of the skyrmion Hall effect, we measure  the value of $Y^{*}_{\mbox{\tiny{S}}}$ that represents the average distance perpendicular to the longest axis of the nanotrack in which the skyrmion is transported inside the region with repulsive edges.  Figures \ref{fig:A_edges}(b),\ref{fig:DM_edges}(b) and \ref{fig:K_edges}(b) show the value of $Y^{*}_{\mbox{\tiny{S}}}$ as a function of the spatial variation of the magnetic property. Such behavior is shown for different widths of the repulsive edges. When analyzing these results, we observed that the  skyrmion Hall effect can be suppressed by adjusting the spatial variation of the magnetic property and the repulsive edge width simultaneously.
A few remarks are in order. At lower values of the local reduction of the strength Dzyaloshinskii-Moriya interaction $D''/D \to 1$, the repulsive force is weak, the skyrmion can overcome the potential barrier and stay inside the repulsive edge. On the other hand, at higher values of the local reduction of the strength Dzyaloshinskii-Moriya interaction $D''/D \to 0$, the repulsive force is too strong, so that the skyrmion vanishes when it enters a narrow region; it disappear collapsing to the ferromagnetic state. That is the reason why there are missing points in Fig. \ref{fig:DM_edges}(b). One can see that the width of repulsive edges is a crucial parameter to transport the skyrmions  inside the center region of the nanotrack.

\begin{figure}[htb!]
\centering
	\includegraphics[width=7.4cm]{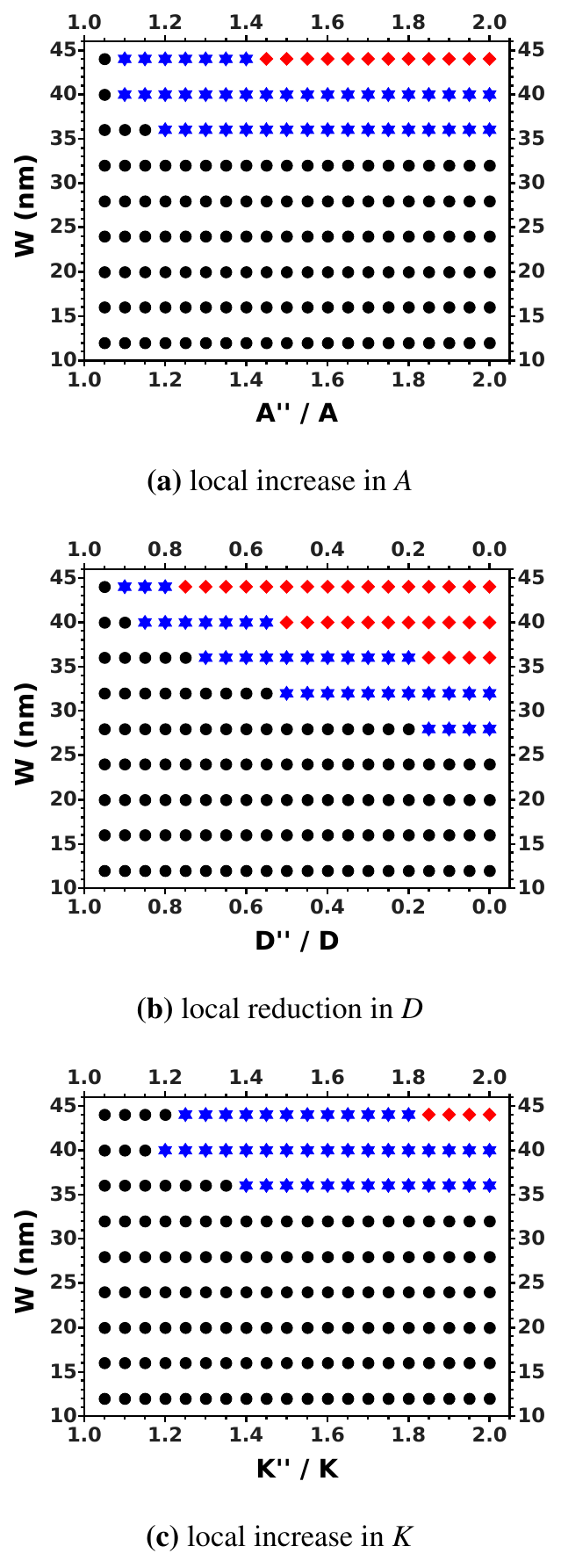}
\caption{(Color online). Diagram for the suppression of the skyrmion Hall effect when using a nanotrack with repulsive edges. Blue stars describe simulation results in which the suppression of the skyrmion Hall effect was observed. Black circles represent the skyrmion transport with Hall effect. For red diamonds the skyrmion disappears, collapsing to the ferromagnetic state.}
\label{fig:Edges_Diagram}
\end{figure}


\begin{figure}[htb!]
\centering
	\includegraphics[width=8.3cm]{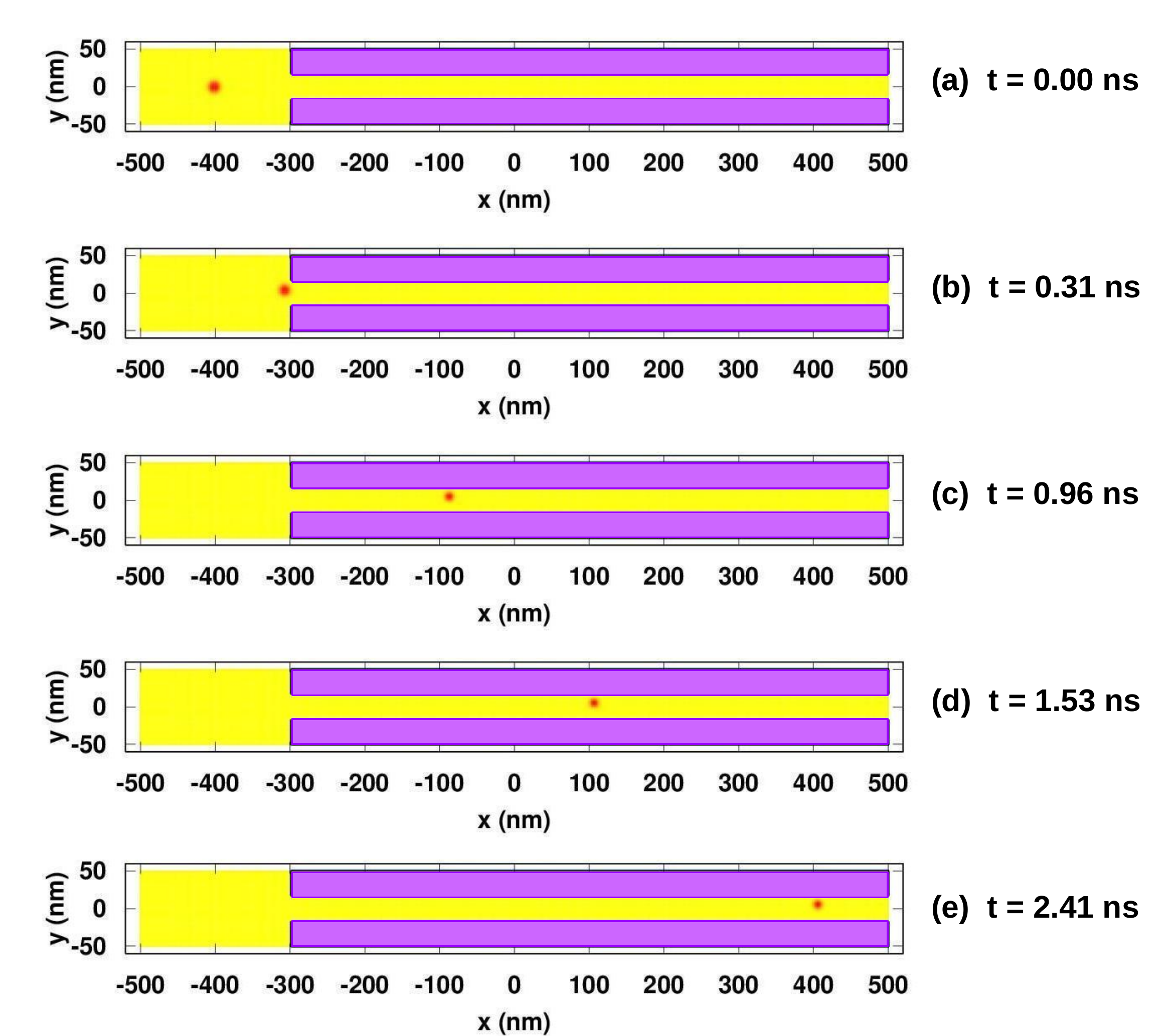}
\caption{(Color online). Suppression of the skyrmion Hall effect. Successive snapshots show the skyrmion being transported inside the center region of the nanotrack with repulsive edges. This micromagnetic simulation has been performed using $D\:''/D =0.50$ and $W=36\:\textrm{nm}$. In this nanotrack magnetically modified, we measure the value of $Y^{*}_{\mbox{\tiny{S}}}=5.6\: \textrm{nm}$.}
\label{fig:Frames_Edges}
\end{figure}


In order to provide a background for experimental studies on the the suppression of the skyrmion Hall effect, we elaborate event diagrams for the results of the micromagnetic simulations. In these diagrams, we have assumed that the skyrmion Hall effect is accurately suppressed as the skyrmion moves around the longest axis of the nanotrack with a tolerance interval of $Y^{*}_{\mbox{\tiny{S}}}\leq R_{\mbox{\tiny{S}}}$. As previously mentioned, the initial radius of the skyrmion is $R_{\mbox{\tiny{S}}} =8.3\: \textrm{nm}$ in our simulations. From the diagrams of the Fig. \ref{fig:Edges_Diagram}, one can observe that the skyrmion Hall effect is suppressed in nanotracks with repulsive edges only for wider edges (blue stars). In the Figs. \ref{fig:Edges_Diagram}(a), \ref{fig:Edges_Diagram}(b) and \ref{fig:Edges_Diagram}(c) the magnitude of the repulsive force with respect to the spatial variation of the magnetic property increases from the left to the right. As the repulsive force is very strong and the skyrmion is forced to cross a very narrow region, it suffers a drastic reduction in its radius and  disappears (red diamonds). Once this mechanism works as a sink for skyrmions, it can be useful to delete information in spintronic devices. 
As expected, if the repulsive force were very weak, the skyrmion transport occurs with considerable Hall effect (black circles), that is, $Y^{*}_{\mbox{\tiny{S}}}> R_{\mbox{\tiny{S}}}$. An example in which the skyrmion Hall effect was suppressed due to the suitable combination of the parameters of repulsive edges is shown in Fig. \ref{fig:Frames_Edges}.


\subsection{\label{sec:center}Nanotracks with an attractive strip}

\begin{figure}[htb!]
\centering
	\includegraphics[width=7.7cm]{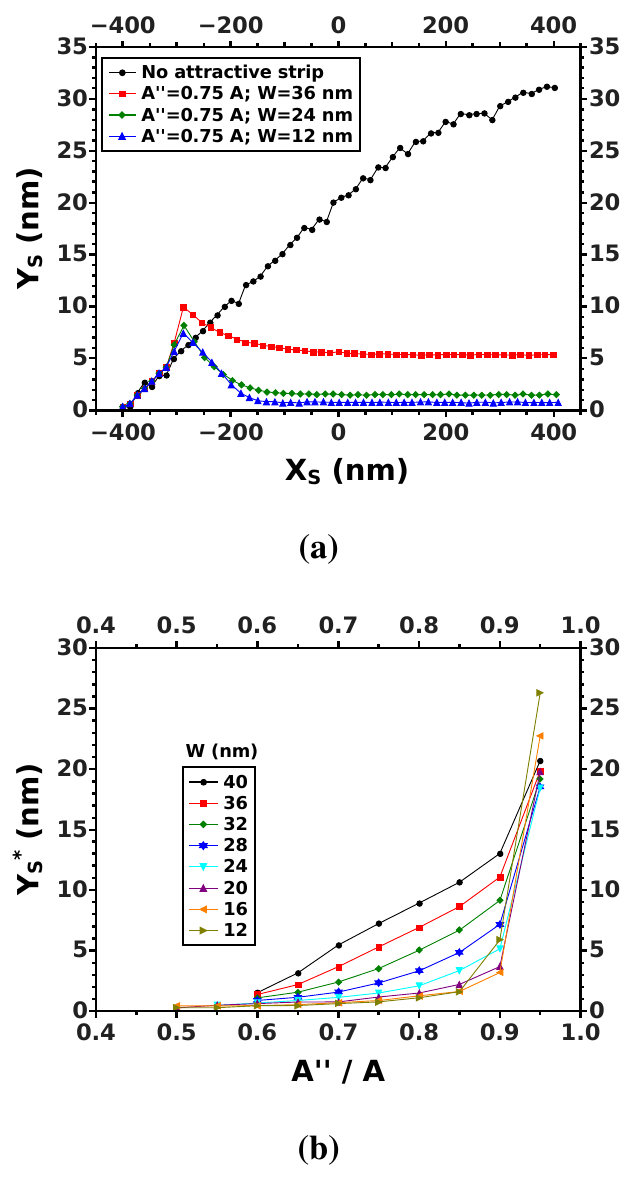}
\caption{(Color online). Spatial variations in the exchange stiffness constant: Type $A$ magnetic defects along the attractive strip of width $W$. Figure (a) shows the skyrmion trajectory in nanotracks with different widths of the attractive strip, presenting a local reduction of 25\% in $A$. Figure (b) shows simulation results for $Y^{*}_{\mbox{\tiny{S}}}$, the variable that quantifies the suppression of the skyrmion Hall effect.}
\label{fig:A_center}
\end{figure}

\begin{figure}[t!]
\centering
	\includegraphics[width=7.7cm]{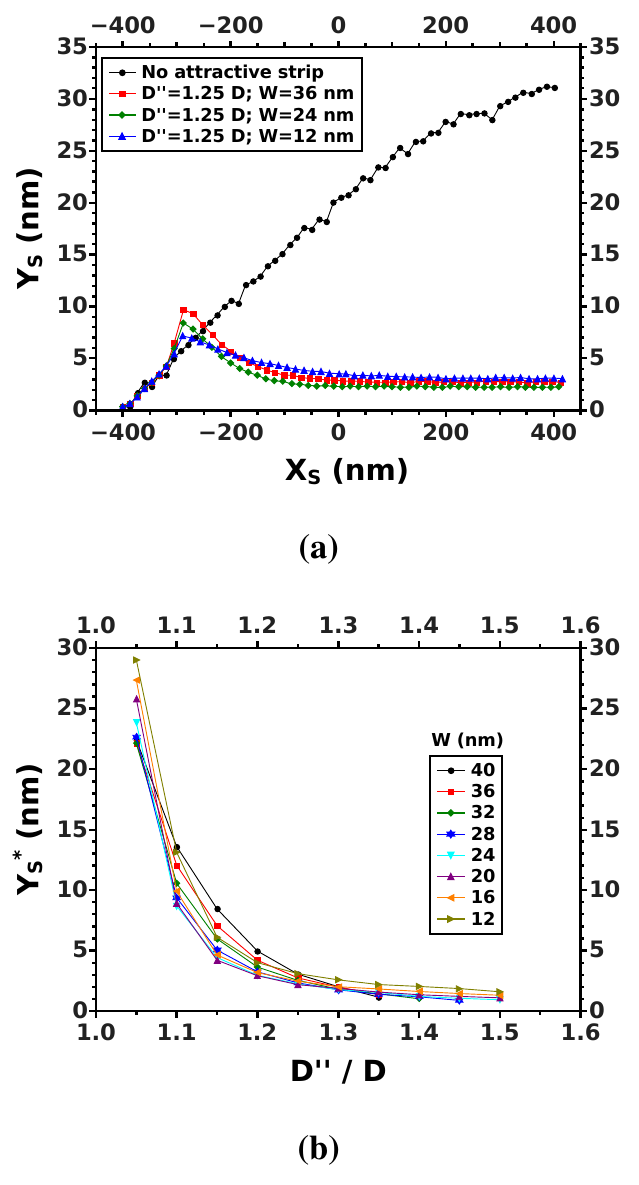}
\caption{(Color online). Spatial variations in the constant of the Dzyaloshinskii-Moriya interaction: Type $D$ magnetic defects along the attractive strip of width $W$. Figure (a) shows the skyrmion trajectory in nanotracks with different widths of the attractive strip, presenting a local increase of 25\% in $D$. Figure (b) shows simulation results for $Y^{*}_{\mbox{\tiny{S}}}$, the variable that quantifies the suppression of the skyrmion Hall effect.}
\label{fig:DM_center}
\end{figure}

\begin{figure}[t!]
\centering
	\includegraphics[width=7.7cm]{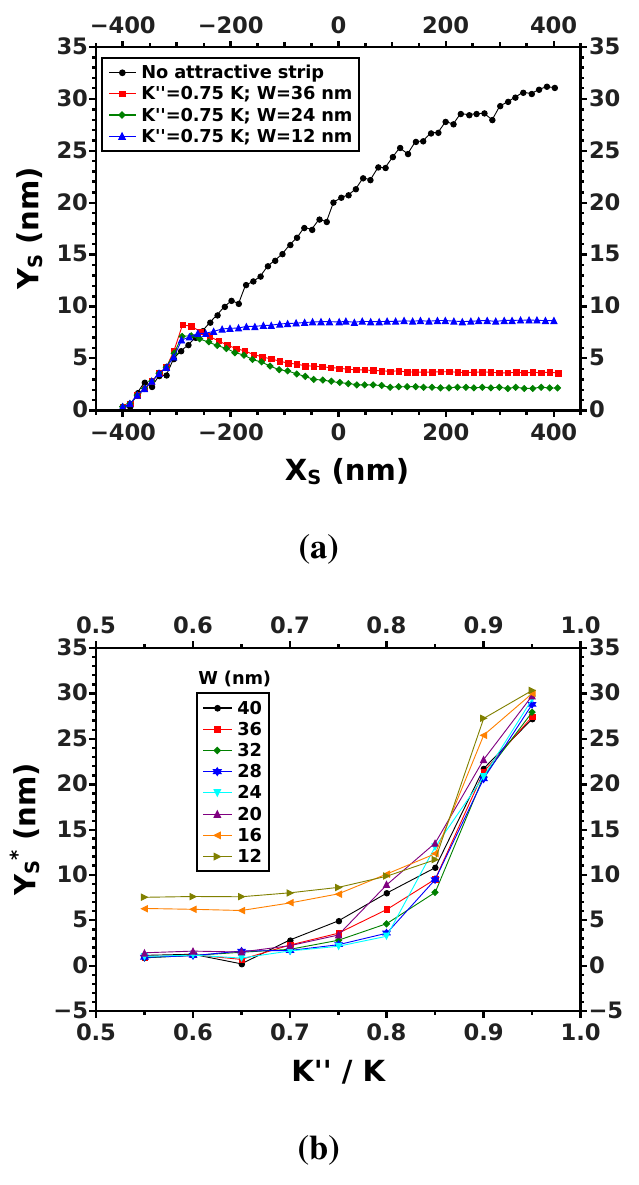}
\caption{(Color online). Spatial variations in the perpendicular magnetic anisotropy: Type $K$ magnetic defects along the attractive strip of width $W$. Figure (a) shows the skyrmion trajectory in nanotracks with different widths of the attractive strip, presenting a local reduction of 25\% in $K$. Figure (b) shows simulation results for $Y^{*}_{\mbox{\tiny{S}}}$, the variable that quantifies the suppression of the skyrmion Hall effect.}
\label{fig:K_center}
\end{figure}

 
In this section, we investigate the suppression of the skyrmion Hall effect in nanotracks with an attractive strip, see Fig. \ref{fig:Schematic}(c). The effect of the attractive strip on the skyrmion trajectory is shown in Figs. \ref{fig:A_center}(a),\ref{fig:DM_center}(a) and \ref{fig:K_center}(a). As a reference, we plot the skyrmion trajectory in the corresponding nanotrack without the attractive strip. Considering our previous results, see Figs. \ref{fig:A_edges}(a),\ref{fig:DM_edges}(a) and \ref{fig:K_edges}(a), one can see that nanotracks with an attractive longitudinal strip are more effective than nanotracks with repulsive edges to suppress the skyrmion Hall effect. It can be inferred from the rapid decay of the value of $Y^{*}_{\mbox{\tiny{S}}}$, the variable that quantifies the suppression of the skyrmion Hall effect; see Figs. \ref{fig:A_center}(b),\ref{fig:DM_center}(b) and \ref{fig:K_center}(b). When analyzing these results, we observed that the  skyrmion Hall effect can be suppressed by adjusting the spatial variation of the magnetic property and the attractive strip width simultaneously.

\begin{figure}[htb!]
\centering
	\includegraphics[width=7.4cm]{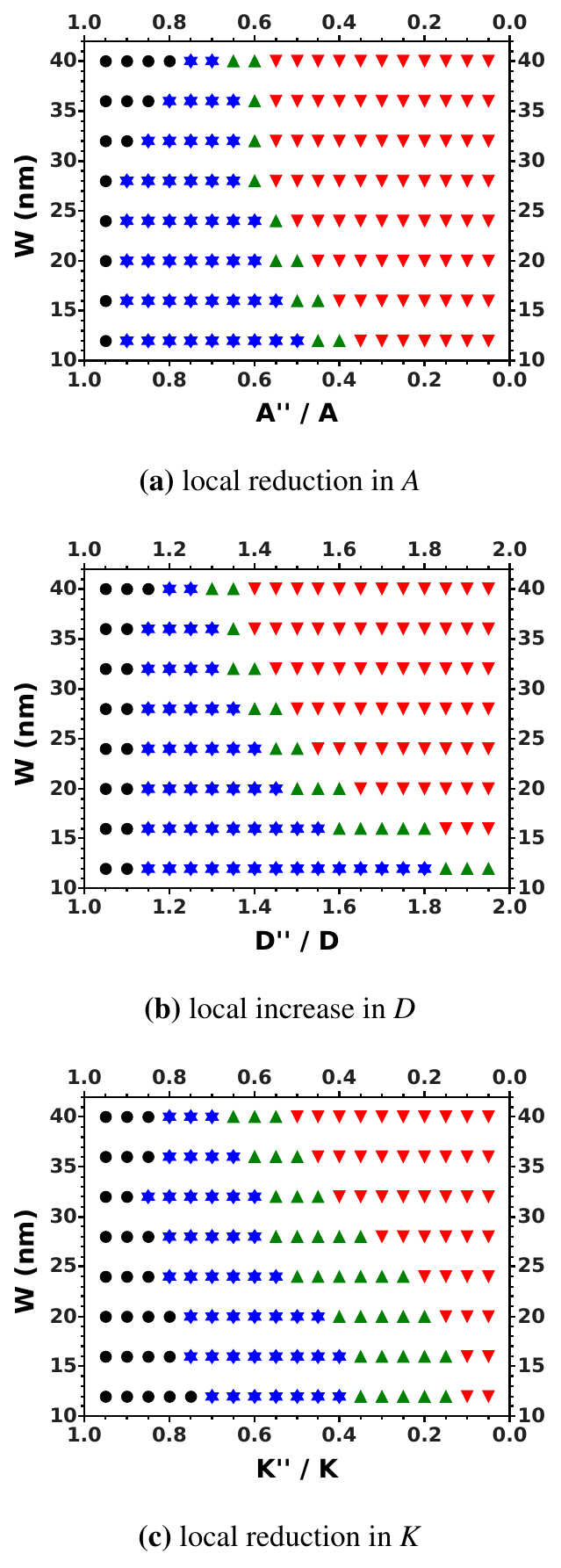}
\caption{(Color online). Diagram for the suppression of the skyrmion Hall effect when using a nanotrack with an attractive longitudinal strip. Black circles represent simulation results in which the skyrmion transport with Hall effect was observed. Blue stars describe the suppression of the skyrmion Hall effect. Green up triangle represent the movement of a deformed skyrmion. Red down triangle, the skyrmion disappears, collapsing to a worm-like magnetic domain.}
\label{fig:Center_Diagram}
\end{figure}

\begin{figure}[htb!]
\centering
	\includegraphics[width=8.3cm]{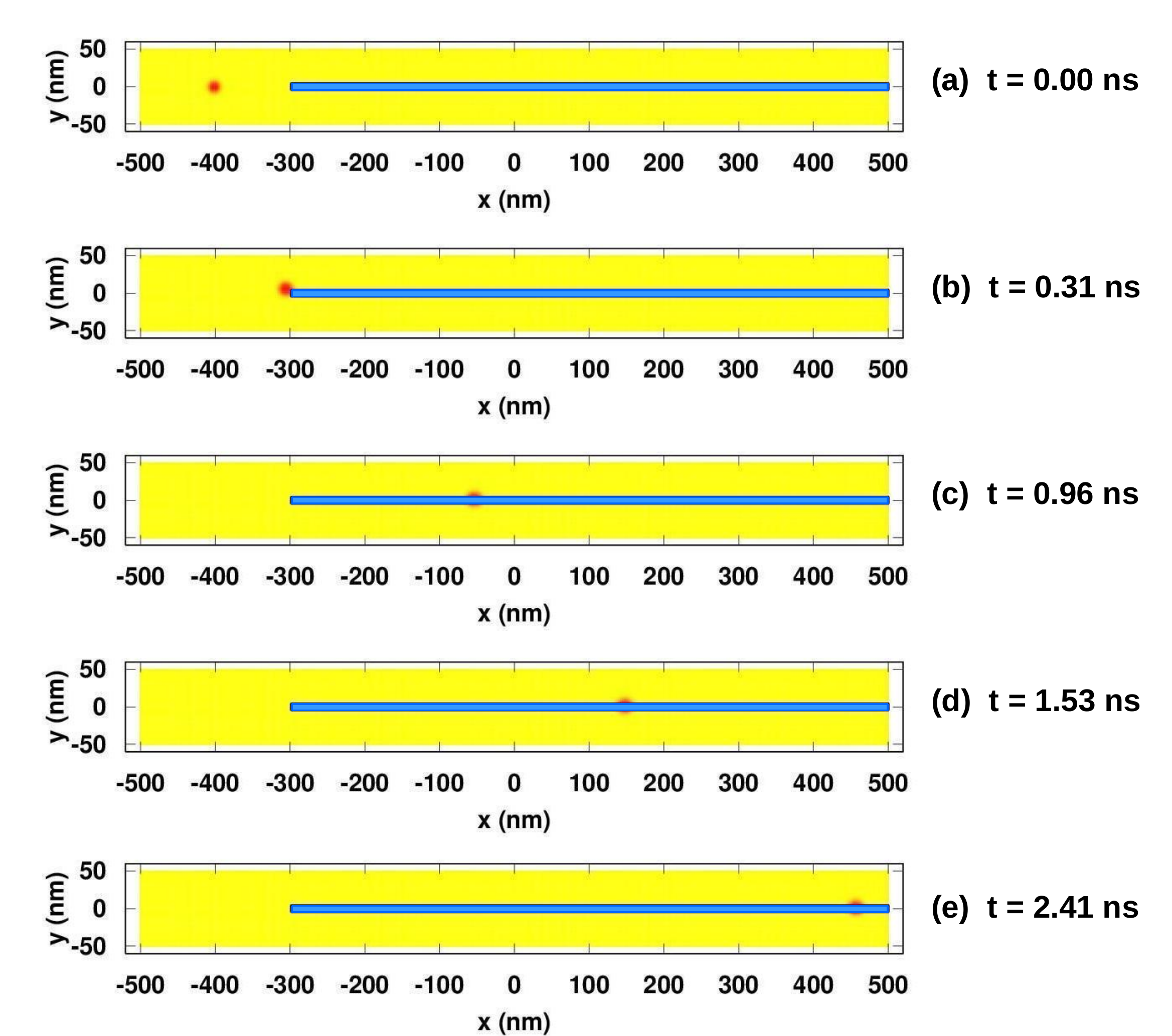}
\caption{(Color online). Suppression of the skyrmion Hall effect. Successive snapshots show the skyrmion being transported inside the center region of the nanotrack with an attractive strip. This micromagnetic simulation has been performed using $D\:''/D =1.50$ and $W=12\:\textrm{nm}$. In this nanotrack magnetically modified, we measure the value of $Y^{*}_{\mbox{\tiny{S}}}=1.6\: \textrm{nm}$.}
\label{fig:Frames_Center}
\end{figure}

Fig. \ref{fig:Center_Diagram} shows that the suppression of the skyrmion Hall effect occurs in a wide range of the attractive strip parameters (blue stars). It is important to mention that in the Figs. \ref{fig:Center_Diagram}(a), \ref{fig:Center_Diagram}(b) and \ref{fig:Center_Diagram}(c) the magnitude of the attractive force with respect to the spatial variation of the magnetic property increases from the left to the right. When the pinning strength is too weak, the skyrmion Hall effect can not be suppressed (black circles); an expected result. On the other hand, when the is pinning strength is strong the skyrmion diameter enlarges considerably, and a flatten particle can be transported inside the attractive strip. Green up triangle represent simulation results in which the movement of a deformed skyrmion was observed. If the pinning strength is too strong, the deformed skyrmion becomes a worm domain inside the attractive longitudinal strip (red down triangle). An example in which the skyrmion Hall effect was suppressed due to the suitable combination of the parameters of the attractive strip is shown in Fig. \ref{fig:Frames_Center}.

\newpage

The results presented here are trustworthy, because our micromagnetic code reproduces and extends the predictions of other groups. In particular, the edge repulsion that originates from a local increase in the perpendicular anisotropy~\cite{Scientific_Reports_7_45330_2017} or from a local reduction of the Dzyaloshinskii-Moriya constant~\cite{PhysRevB_95_144401_2017}. Furthermore, we provide other means of confining the skyrmion along the center region of the nanotrack, that is, the suppresion of the skyrmion Hall effect via spatial modification  of the exchange stiffness constant. Besides the edge repulsion, experimental studies can explore the usage of nanotracks with an attractive   strip~\cite{IEEE_51_1500204_2015}, being that a route is presented in the Fig. \ref{fig:Center_Diagram}. Very sharp variations on the material parameters can not be easy to get and to control experimentally. However, we present many ways of suppressing the skyrmion Hall effect. Although the results presented here are for a very simple distribution of magnetic defects into a ferromagnetic nanotrack, we believe their consequences can be planned and extended to future experimental works.

From the graphs of the skyrmion trajectory, one can see that there is a noise mainly on the transverse coordinate ($Y_{\mbox{\tiny{S}}}$). Once this noise scales with the size of the micromagnetic mesh (2 nm), we believe it is linked to the resolution of the algorithm that was used to track the skyrmion position. Among the algorithms that we have tested~\cite{mendona2019frogs}, we found out the algorithm known as Frogs Method was the most accurate. For this reason, it was used to quantify the suppression of the skyrmion Hall effect in this work.

\section{Conclusion}
\label{Concl}

In summary, the skyrmion Hall effect can be an issue for spintronic devices based on the skyrmion transport in a ferromagnetic nanotrack. Nowadays, the focus of research has been on skyrmions in antiferromagnetic media, because there is no skyrmion Hall effect. Due to the scarcity of antiferromagnetic materials in comparison with ferromagnetic materials~\cite{PhysRevLett_116_147203_2016}, and once ferromagnetic skyrmions have already been stabilized at room temperature~\cite{Science_349_283_2015,ApplPhysLett_106_242404_2015,Nat_Mater_15_501_2016,NatureNanoTechnology_11_444_2016,NatureNanoTechnology_11_449_2016,
ApplPhysLett_111_202403_2017,ApplPhysLett_112_132405_2018}, we decide to investigate the possibility of suppressing the skyrmion Hall effect. In this work, we present 6 ways to avoid the skyrmion accumulation at the nanowire edges or even the loss of information in the case in which the skyrmions are expelled from the nanowire. Our predictions to suppress the skyrmion Hall effect are summarized in the Figures \ref{fig:Edges_Diagram} and \ref{fig:Center_Diagram}. In a previous work~\cite{JMagnMagnMater_480_171_185_2019}, our team showed that spatial variations on the material parameters of the ferromagnetic nanotrack (exchange stiffness, saturation magnetization, magnetocrystalline anisotropy and Dzyaloshinskii-Moriya constants) can work as traps for pinning and scattering magnetic skyrmions. In that work, we showed that the skyrmion was attracted to a trap characterized by a local increase in the Dzyaloshinskii-Moriya constant $(D''\:>\:D)$, whereas the skyrmion was repelled from a trap characterized by a local reduction in the Dzyaloshinskii-Moriya constant $(D''\:<\:D) $. Here, we have investigated the suppression of the skyrmion Hall effect in  nanotracks with their magnetic properties strategically modified along the longitudinal strips. In particular, we study two categories of nanotracks modified magnetically. One of them, repulsive edges have been inserted in the nanotrack and, in the other, an attractive strip has been placed exactly on the longest axis of the nanotrack. The skyrmion motion driven by a spin-polarized current has been studied through micromagnetic simulations. Our results indicate that the skyrmion motion can be along the center region of the nanotrack, provided their magnetic properties be strategically modified. Results here presented provide a background for experimental studies. We hope this study encourages experimental works to investigate the suppression of the skyrmion Hall effect in nanotracks modified strategically via engineering of magnetic properties. Although we have not considered Type $M_{\mbox{\tiny{S}}}$ magnetic defects (those characterized only by spatial variations in the saturation magnetization) in this paper, experimental works can investigate the suppression of the skyrmion Hall effect by the usage of nanotracks with repulsive edges $(M_{\mbox{\tiny{S}}}''< M_{\mbox{\tiny{S}}})$ and/or nanotracks with an attractive strip $(M_{\mbox{\tiny{S}}}''> M_{\mbox{\tiny{S}}})$. Such predictions stem from our previous work~\cite{JMagnMagnMater_480_171_185_2019}. Regardless of those results which took into account  the effect of the full dipolar coupling, these predictions can be also derived from the equation (\ref{eq:K_eff}); the  effective uniaxial anisotropy. As the $M_{\mbox{\tiny{S}}}$ parameter is linked to the dipolar coupling (the weakest magnetic interaction), we believe that sharp spatial variations in $M_{\mbox{\tiny{S}}}$ are needed. Thus, the attractive or repulsive interactions become significant and the dynamics of the ferromagnetic skyrmion can be manipulated. While this paper was being reviewed, we found that other groups have recently proposed alternative means of suppressing the skyrmion Hall effect. For example, Ref.~\cite{PhysRevB_99_020405_2019} reports the suppression of the skyrmion Hall effect by tuning the parameters of torque exerted on the magnetization by a spin current, which is generated by spin-Hall effect~\cite{Scientific_Reports_4_6784_2014}. Another possibility that has been explored is the usage of ferromagnetic bilayer coupled antiferromagnetically~\cite{Nat_Commun_7_10293_2016}. The skyrmion Hall effect can be also suppressed with vortex capping and the aid of RKKY coupling~\cite{JMagnMagnMater_455_39_2018}. In a future work, we intend to contribute to the development of the Antiferromagnetic Spintronics.

\section*{Acknowledgments}
\label{Ackno}
The authors would like to thank CAPES, CNPq, FAPEMIG and FINEP (Brazilian Agencies) for the support. Numerical works were done at the Laborat\'orio de Simula\c c\~ao Computacional do Departamento de F\'isica da UFJF.
%


%
\end{document}